\documentclass[journal]{IEEEtran}
\usepackage{amsmath,amsfonts}
\usepackage{algorithmic}
\usepackage{algorithm}
\usepackage{array}
\usepackage[caption=false,font=footnotesize,labelfont=rm,textfont=rm]{subfig}
\usepackage{textcomp}
\usepackage{stfloats}
\usepackage{url}
\usepackage{verbatim}
\usepackage{graphicx}
\usepackage{cite}
\usepackage{xcolor}
\newcommand{\imagewidth}{0.11\textwidth}
\newcommand{\imagespace}{-0.3cm}
\newcommand{\sectionspace}{-0.3cm}

\begin{document}

\title{Prompt-Assisted Semantic Interference Cancellation on Moderate Interference Channels}

\author{Zian Meng, Qiang Li,~\IEEEmembership{Member,~IEEE}, Ashish Pandharipande,~\IEEEmembership{Senior Member,~IEEE}, \\and Xiaohu Ge,~\IEEEmembership{Senior Member,~IEEE}

\vspace{-0.1cm}

\thanks{Zian Meng, Qiang Li (Corresponding author), and Xiaohu Ge are with Huazhong University of Science and Technology, 430074 P. R. China (emails: sarfflow@hust.edu.cn, qli\_patrick@hust.edu.cn, xhge@mail.hust.edu.cn). A. Pandharipande is with NXP Semiconductors, 5656 AE Eindhoven, The Netherlands (email: ashish.pandharipande@nxp.com).}
\thanks{This work is partially supported by the Hubei Provincial Key R\&D Program under grant 2022EHB014.}}

\maketitle

\begin{abstract}

The performance of conventional interference management strategies degrades when interference power is comparable to signal power. We consider a new perspective on interference management using semantic communication. Specifically, a multi-user semantic communication system is considered on moderate interference channels (ICs), for which a novel framework of deep learning-based prompt-assisted semantic interference cancellation (DeepPASIC) is proposed. Each transmitted signal is partitioned into common and private parts. The common parts of different users are transmitted simultaneously in a shared medium, resulting in superposition. The private part, on the other hand, serves as a prompt to assist in canceling the interference suffered by the common part at the semantic level. Simulation results demonstrate that the proposed DeepPASIC outperforms conventional interference management strategies under moderate interference conditions.

\end{abstract}


\vspace{\sectionspace}
\section{Introduction}
The explosive growth in the number of devices and data traffic in wireless networks has resulted in higher demands for network coverage and throughput \cite{shen2022novel}. However, network density and resource reuse have inevitably increased interference within cells, between cells, and across the networks, making interference a bottleneck that constrains the system performance \cite{ifManagement}. Semantic communication, which focuses on transmitting the meaning of the message, is a promising solution to this challenge due to its potential to intelligently extract the desired content from the interfered signal \cite{yang2022semantic}.

%The interference channel (IC) describes a classic communication model where multiple non-cooperative transmitters communicate with their intended receivers through a shared medium, such as a pair of base stations and terminal devices near a cell boundary. 

Interference channels (ICs) have been a longstanding challenge in Shannon's Information Theory. Traditional heuristic interference management schemes, such as orthogonal channel allocation, treating interference as noise (TIN), and successive interference cancellation (SIC), do not ensure optimal performance, particularly in moderate interference regimes \cite{prsma}. More sophisticated approaches, including the Han-Kobayashi (HK) scheme \cite{han1} and rate-splitting multiple access (RSMA) \cite{srsma}, offer a more refined solution by dividing the transmitted message into common part (portion transmitted simultaneously) and private part (portion transmitted orthogonally). This division enables each receiver to partially decode information intended for other transmitters, thereby achieving enhanced capacities across a range of interference levels. However, when the signal power is comparable to the interference power, the capacity gains offered by rate splitting are limited to those achievable with orthogonal schemes \cite{etkin2008gaussian}. Moreover, rate-splitting schemes necessitate joint decoding of the common and private parts, which poses a significant challenge in terms of designing and implementing complex decoding strategies \cite{hank}.

% Traditionally, the fundamental concept of interference separation involves isolating the non-orthogonal resource domains of different users, as exemplified in technologies such as power-domain non-orthogonal multiple access (PD-NOMA) and code-domain NOMA (CD-NOMA) \cite{codingDomain}. 

Recently, advancements in semantic communication have demonstrated that interference separation can be achieved at the semantic level within identical conventional resource domains, such as time, frequency, and space. For example, Zhang et al. \cite{ifsnoma} presented a semantic-aware interference cancellation scheme for downlink non-orthogonal multiple access (NOMA). Tang et al. \cite{tanghnoma} investigated a semantic-assisted hybrid NOMA with heterogeneous users utilizing bit and semantic communication. Ma et al. \cite{sfdma} proposed an interference management scheme in the semantic feature domain for ICs. Lin et al. \cite{semanticic} leveraged semantic prior knowledge as side information to enhance traditional channel coding. Despite these advances, existing schemes often require cooperation between transmitters or restrict each transceiver pair to transmit information within a specific semantic domain, potentially limiting the flexibility of the transmitted content. Inspired by advancements in speech and image separation \cite{ zou2020deep}, it is possible to separate superposed semantic features. However, superposition inevitably leads to information loss, referred to as ambiguity, such as the label permutation problem. To address this issue, rate splitting is a promising approach that helps resolve the ambiguity. Specifically, information susceptible to superposition is allocated to the private parts, which then serve as prompts to assist in separating the overlapping common part through feature fusion techniques such as those used in \cite{tang2023cooperative}.

In this letter, the challenges of interference management in ICs are revisited from the perspective of semantic communication. A new DL-based prompt-assisted semantic interference cancellation (DeepPASIC) framework is proposed, enabling insights into interference separation at the semantic level. The major contributions are summarized as follows:
\begin{itemize}
    \item A multi-user semantic communication system is established on ICs, based on which a new interference management framework named DeepPASIC is proposed. To be specific, multiple transmitter-receiver pairs with identical codecs communicate over a Gaussian IC, where prompt information is utilized to reduce ambiguity and improve transmission reliability.
    
    \item The transmitted message from each user is partitioned into common and private parts. The common part is susceptible to interference from other users, whereas the private part is transmitted orthogonally. With the aid of the private part that serves as a prompt, the receiver employs a generative adversarial network (GAN) \cite{CycleGAN2017} to eliminate the interference suffered by the common part.
    
    \item Simulation results demonstrate the effectiveness of the proposed DeepPASIC for image transmission and reconstruction. Specifically, significant performance improvements in peak signal-to-noise ratio (PSNR) are achieved in the moderate interference regime, as compared to conventional interference management schemes.
\end{itemize}

\vspace{\sectionspace}
\section{System Model}
In this section, the system model of multi-user semantic communication on ICs is established, where the transmitted message from each user is partitioned into two parts that are respectively delivered over two stages.

\vspace{\sectionspace}
\subsection{Interference Channel}
We focus on $K$-user ICs with additive white Gaussian noise (AWGN). Given the shared communication medium, the receiver for each user receives a superposition of the transmitted signals from all transmitters, along with the AWGN.

Denoting the transmitted signal of user $k$ as $\mathbf{x}_k \in \mathbb{C}^M$, for $k = 1, 2, \ldots, K$, the corresponding received signal matrix can be expressed as
\begin{equation}
\mathbf{Y} = \mathbf{HX} + \mathbf{N},
\end{equation}
where $\mathbf{Y} = [\mathbf{y}_1, \mathbf{y}_2, \ldots, \mathbf{y}_K]^T \in \mathbb{C}^{K\times M}$ is the received signal matrix, $\mathbf{X} = [\mathbf{x}_1, \mathbf{x}_2, \ldots, \mathbf{x}_K]^T \in \mathbb{C}^{K\times M}$ is the transmitted signal matrix, and $\mathbf{N} = [\mathbf{n}_1, \mathbf{n}_2, \ldots, \mathbf{n}_K]^T \in \mathbb{C}^{K\times M}$ is the noise matrix. The channel state information matrix is denoted by $\mathbf{H} = [h_{ij}]_{K\times K}$, where $h_{ij}$ represents the channel coefficient from transmitter $j$ to receiver $i$. The noise vector $\mathbf{n}_{k}$ is characterized by i.i.d. complex AWGN with a power of $\sigma^{2}$, denoted as $\mathbf{n}_{k}\sim\mathcal{CN}(0,\sigma^2\mathbf{I})$.

% \begin{figure}
%     \centering
%     \subfloat[$K$-user semantic communication on IC.]{
%         \includegraphics[width=0.40\textwidth]{transmit strategy2 1.pdf}
%     }\label{fig:IC system model}\\
%     \subfloat[Two-stage transmission strategy.]{
%         \includegraphics[width=0.40\textwidth]{transmit strategy3.pdf}
%     }\label{fig:two-stage transmission}
%     \caption{System model and two-stage transmission strategy.}
%     \label{fig:system model and stage}
% \end{figure}

\vspace{\sectionspace}
\subsection{Two-Stage Transmission}

\begin{figure*}[!t]
    \centering
    \includegraphics[width=0.95\textwidth]{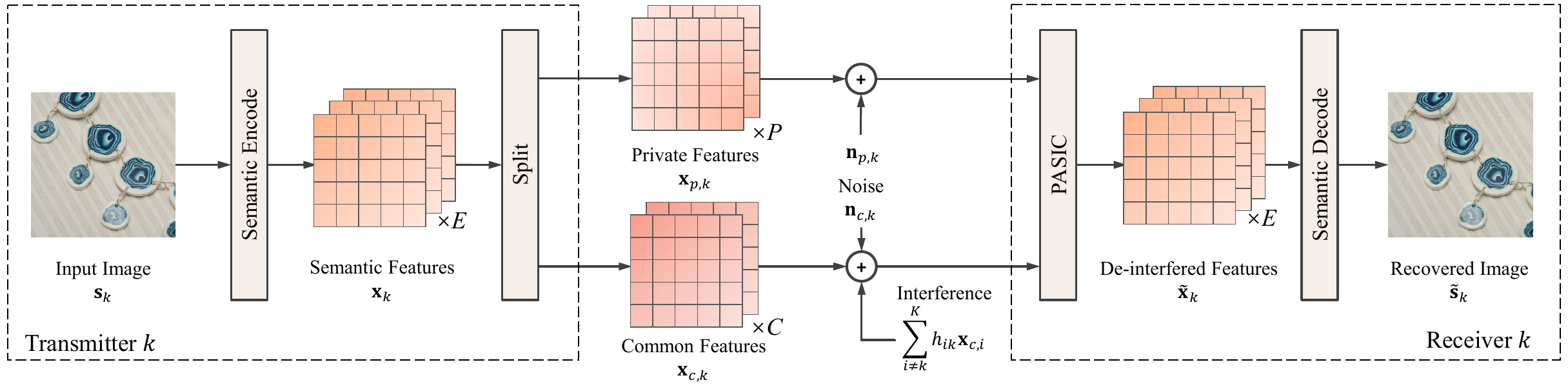}
    \caption{The framework and data pipeline of proposed DeepPASIC at a typical user $k$.}
    \vspace{\imagespace}
    \label{fig:pasicfw}
\end{figure*}

The proposed DeepPASIC framework is presented in Fig. \ref{fig:pasicfw}. At the $k$-th transmitter, the source information $\mathbf{s}_k$ is passed through a trainable semantic encoder $E_{\phi}$ with trainable parameters $\phi$, to obtain the transmitted signal
\begin{equation}
\mathbf{x}_{k} = E_{\phi}(\mathbf{s}_{k}).
\end{equation}
Then, $\mathbf{x}_{k}\in \mathbb{C}^M$ is split into a common part $\mathbf{x}_{c,k} \in \mathbb{C}^{U}$ and a private part $\mathbf{x}_{p,k} \in \mathbb{C}^{V}$, $U+V=M$. The common part should have an easily separable structure, while the private part should contain information susceptible to interference.

%\begin{equation}
%\left\{ \mathbf{x}_{c,k}, \mathbf{x}_{p,k} \right\}=\text{Split}(\mathbf{x}_{k}).
%\end{equation}

\begin{figure}[t]
\centering
\includegraphics[width=0.40\textwidth]{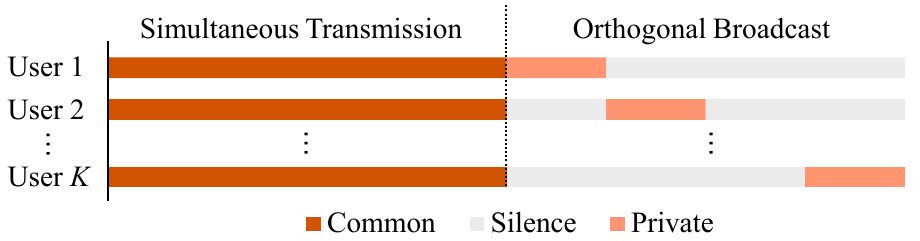}
\caption{Two-stage transmission strategy.}
\vspace{\imagespace}
\label{fig:system model and stage}
\end{figure}

The transmission process is divided into the simultaneous transmission (ST) stage and the orthogonal broadcast (OB) stage, as shown in Fig. \ref{fig:system model and stage}. In the ST stage, the common parts of all users are transmitted in the same resource block, along with the corresponding received signal matrix
\begin{equation}
\mathbf{Y}_{c} = \mathbf{H}\mathbf{X}_{c} + \mathbf{N}_{c}, \label{eq:common channel}
\end{equation} 
where $\mathbf{Y}_c = [\mathbf{y}_{c,1}, \mathbf{y}_{c,2}, ..., \mathbf{y}_{c,K}]^{T} \in \mathbb{C}^{K\times U}$ is the received common signal matrix, $\mathbf{X}_c = [\mathbf{x}_{c,1}, \mathbf{x}_{c,2}, ..., \mathbf{x}_{c,K}]^{T} \in \mathbb{C}^{K\times U}$ is the transmitted common signal matrix, and $\mathbf{N}_c = [\mathbf{n}_{c,1}, \mathbf{n}_{c,2}, ..., \mathbf{n}_{c,K}]^{T} \in \mathbb{C}^{K\times U}$ is the noise matrix.

In the OB stage, the private part of each user is transmitted orthogonally using $K$ isolated resource blocks through time division or frequency division. To be specific, in the $j$-th block where $j=1,2,\dots, K$, only the $j$-th transmitter is activated to send its message while others are deactivated. For all receivers in the $j$-th block, we have the corresponding received signal matrix
\begin{equation}
\mathbf{Y}_{p,j} = \mathbf{H} \mathbf{X}_{p,j} + \mathbf{N}_{p,j},
\label{eq:private channel}
\end{equation} 
where $\mathbf{Y}_{p,j} = [\mathbf{y}_{p,j,1}, \mathbf{y}_{p,j,2}, ..., \mathbf{y}_{p,j,K}]^{T} \in \mathbb{C}^{K\times V}$ is the received prompt signal matrix, $\mathbf{X}_{p,j} = [0, 0, ..., \mathbf{x}_{p,j}, ..., 0]^{T} \in \mathbb{C}^{K\times V}$ is the transmitted prompt signal matrix, and $\mathbf{N}_p = [\mathbf{n}_{p,j,1}, \mathbf{n}_{p,j,2}, ..., \mathbf{n}_{p,j,K}]^{T} \in \mathbb{C}^{K\times V}$ is the noise matrix.
The total orthogonal resources consumed by the two-stage transmission are represented by $S = U + KV$. Given the total orthogonal resource $S$ and semantic encoding length $M$, the length of the private part can be calculated as
\begin{equation}
    V=\frac{S-M}{K-1}.
\end{equation}

After the two transmission stages, the $k$-th receiver will receive a common signal overlaid by the common parts of all users and $K$ independent private signals from all users. These signals will be jointly processed by the separator $G_{\gamma}$, at which the private signals from each user are used to guide the interference cancellation for the common signal, obtaining the expected signal
\begin{equation}
    \tilde{\mathbf{x}}_{k} = G_{\gamma} \left(\mathbf{y}_{c,k}, \mathbf{y}_{p,1,k}, \mathbf{y}_{p,2,k},...,\mathbf{y}_{p,K,k}\right).
\end{equation}
Note that while the exact separation of superimposed analog signals is generally unachievable, semantic communication does not require such precision, as the same semantic information can be represented by multiple distinct symbols \cite{sit}, allowing for effective meaning transmission even with imperfect signal separation. Finally, a semantic decoder $D_{\theta}$ with trainable parameters $\theta$ attempts to reconstruct the source message from the separated signal
\begin{equation}
    \tilde{\mathbf{s}}_{k}=D_{\theta}(\tilde{\mathbf{x}}_{k}).
\end{equation}

\vspace{\sectionspace}
\section{Proposed DeepPASIC Framework}
As shown in Fig. \ref{fig:pasicfw}, this section introduces the neural network architecture of the proposed DeepPASIC, including the encoder, channel, decoder, and separator. Then, the loss function and training procedure are discussed in detail.

\vspace{\sectionspace}
\subsection{Network Architecture of DeepPASIC}

\subsubsection{Semantic Encoder}

For the $k$-th user where $k=1,2,\dots, K$, the encoder consists of three cascaded residual convolutional blocks, a structure proven to be effective in extracting the semantic features of images \cite{PADC}. The input to the encoder is a colored image $\mathbf{s}_{k}$ with a shape of $[3,H,W]$. After two downsampling operations, the semantic featuremap $\mathbf{x}_{k}=E_{\phi}(\mathbf{s}_{k})$ with a shape of $[E,H/4,W/4]$ is obtained, where $E$ denotes the number of output feature layers of the encoder, and the total length $N=E\times H/4\times W/4$ of the semantic encoding can be determined.

\subsubsection{Channel}
The output of the semantic encoder is then split into private and common parts at the channel layer. To be specific, the first $P$ layers represent the private part with a shape of $[P,H/4,W/4]$, and the remaining layers constitute the common part with a shape of $[C,H/4,W/4]$, where $C=E-P$ denotes the number of common part layers.

On this basis, the common and private parts are respectively transmitted to the receiver through IC, according to the two-stage transmission strategy described in Section II. 

\subsubsection{GAN Separator}
The received signals are concatenated and fed into the GAN separator, at which the interference suffered by the common part is eliminated by utilizing the private parts as prompts. Specifically, at the $k$-th receiver, both the common and private parts are first zero-padded to a uniform shape of $[E, H/4, W/4]$ to facilitate flexible allocation of the common and private parts, then processed as follows:
\begin{equation}
    \mathbf{y}_k = \text{Concat}\left(\mathbf{y}_{c,k} + \mathbf{y}_{p,1,k}, \mathbf{y}_{p,2,k}, ..., \mathbf{y}_{p,K,k}\right),
\end{equation}
where $\text{Concat}(\cdot)$ represents the concatenation operation along the channel dimension, and the shape of the concatenated tensor $\mathbf{y}_k$ is $[K \times E, H/4, W/4]$. Note that the first private part is added to the common part to conserve memory and reduce unnecessary computational overhead. This operation is equivalent to concatenating the common part and the first private part without zero padding. Subsequently, the predicted common parts of all users can be obtained by
\begin{equation}
\{\tilde{\mathbf{x}}_{c,1},\tilde{\mathbf{x}}_{c,2},\dots,\tilde{\mathbf{x}}_{c,K}\} = G_{\gamma} \left(\mathbf{y}_k\right),
\end{equation}
from which the desired $\tilde{\mathbf{x}}_{c,k}$ can be selected. 

\subsubsection{Semantic Decoder}
The decoder, which consists of three residual transposed convolutional blocks, is utilized to reconstruct the original transmitted image from the merge of the separated common part and the corresponding private part:
\begin{equation}
    \tilde{\mathbf{s}}_{k}=D_{\theta}(\tilde{\mathbf{x}}_{c,k} + \mathbf{y}_{p,k,k}).
\end{equation}
By leveraging prior knowledge on the image features and the prompt provided by the private part, the decoder can further eliminate the residual interference in the predicted common part from the separator, thus effectively enhancing the end-to-end image reconstruction quality.

\vspace{\sectionspace}
\subsection{Training Algorithm}
The training algorithm proceeds in a 3-step process: (1) train the autoencoder without interference and separator, (2) train the separator on ICs using the pre-trained autoencoder, and (3) alternately optimize the autoencoder and separator.

\subsubsection{Train the autoencoder}
The semantic encoder and decoder are trained separately to construct the semantic space of the image data initially. The mean squared error (MSE) of pixel values between the original image and the reconstructed image is used as the loss function: 
\begin{align}\label{eq:phase1loss}
    \mathcal{L}_{\phi,\theta}=\mathbb{E}\|\mathbf{s}_k-\tilde{\mathbf{s}}_k\|^2.
\end{align}

\subsubsection{Train the GAN separator}
Given the parameters of the autoencoder trained in the previous step, the separator, composed of a generator and a discriminator, is trained to eliminate interference from the superimposed signal through adversarial training. The generator outputs a separation result $\tilde{\mathbf{x}}_k$ for the $k$-th user, then the discriminator is used to evaluate the separation quality. The loss function is the sum of the binary cross-entropy loss of the discriminator, which can be expressed as:
\begin{align}
    \mathcal{L}_{D}=\frac{1}{K}\sum^K_{k=1}\log D_\mu(\mathbf{x}_k)+\log\left(1-D_\mu\left(\tilde{\mathbf{x}}_k\right)\right),
\end{align}
where $D_\mu$ represents the discriminator with trainable parameters $\mu$. The discriminator is trained to distinguish between real and fake samples. Therefore, it is not active during the inference phase when separating the interference using the generator.
To separate the interference from the superimposed signal, the losses between each user's separation result $\tilde{\mathbf{x}}_k$ and its original signal $\mathbf{x}_k$ are measured in terms of MSE, i.e.,
\begin{equation}
    \mathcal L_{G} = \frac{1}{K}\sum^K_{k=1}\mathbb{E}\|\mathbf{x}_k-\tilde{\mathbf{x}}_k\|^2.
    \label{eq:trainG}
\end{equation}
At the same time, in order to generate results that approach interference-free features, the generator also needs to deceive the discriminator as much as possible. The optimization object can then be expressed as: 
\begin{align}
    \mathcal{G}^*=\arg\min_{\gamma}\max_{\mu}\mathbb{E}\left[\alpha\mathcal L_{G}-\beta\mathcal{L}_{D}\right],
\end{align}
where $\alpha$ and $\beta$ are the hyperparameters to control the balance between the generator and discriminator losses, respectively. 

\subsubsection{Alternating Optimization of the Whole Network}

An alternating optimization strategy is applied to refine both the autoencoder and the separator in an end-to-end manner. At the training iteration $t$, the separator's parameters are fixed at $\mathcal{G}^*(t-1)$, and the autoencoder's parameters $(\phi(t), \theta(t))$ are updated to minimize the reconstruction error, thus obtaining optimal $(\phi^*(t), \theta^*(t))$. Subsequently, keeping the autoencoder parameters at their optimized state, the separator is fine-tuned to further improve the interference separation capability and obtain $\mathcal{G}^{*}(t)$. This iterative process continues until the end-to-end reconstruction loss converges, enhances the overall image restoration quality, and enables the separator to adapt to changes in the semantic coding space.

% By iteratively refining both components, the system achieves a balanced and effective semantic communication framework capable of handling interference while preserving semantic integrity.

\begin{figure*}[htbp]
    \centering
    \begin{tabular}{ccccc}
    \multicolumn{1}{c}{\textbf{Transmitted 1}} & \multicolumn{1}{c}{\textbf{Transmitted 2}} & \multicolumn{1}{c} {\textbf{Interfered}} & \multicolumn{1}{c} {\textbf{Received 1}} & \multicolumn{1}{c}{\textbf{Received 2}} \\
        \includegraphics[width=\imagewidth]{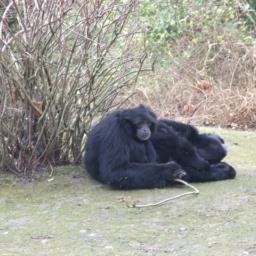} &
        \includegraphics[width=\imagewidth]{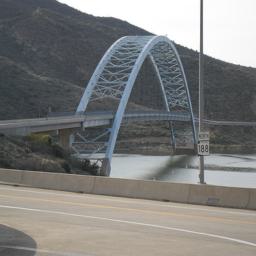} &
        \includegraphics[width=\imagewidth]{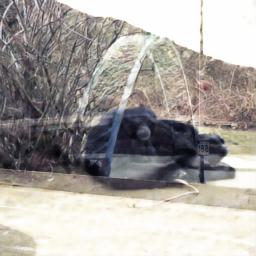} &
        \includegraphics[width=\imagewidth]{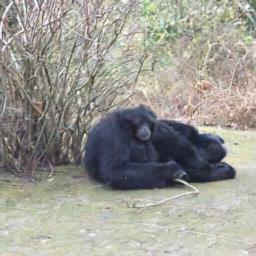} &
        \includegraphics[width=\imagewidth]{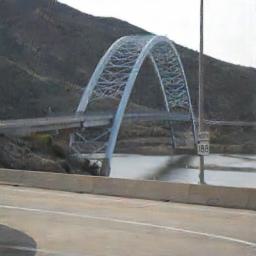} \\
        \includegraphics[width=\imagewidth]{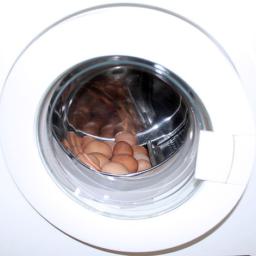} &
        \includegraphics[width=\imagewidth]{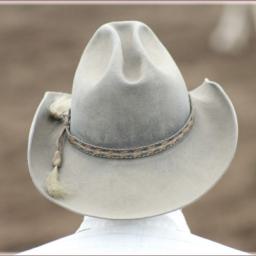} &
        \includegraphics[width=\imagewidth]{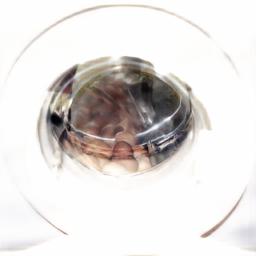} &
        \includegraphics[width=\imagewidth]{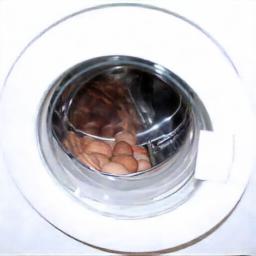} &
        \includegraphics[width=\imagewidth]{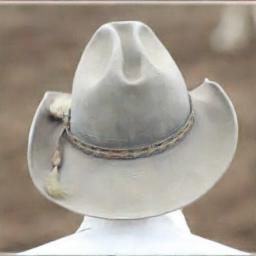} \\
        \text{(a)} & \text{(b)} & \text{(c)} & \text{(d)} & \text{(e)}
    \end{tabular}
    \caption{Some visualized results for two-user DeepPASIC under moderate IC, where $|h_{ij}|=1$, $C=12$, and $P=4$. (a), (b) represent original images encoded and transmitted by the two users, respectively. (c) displays the reconstruction results of both users without interference separation, which are identical due to the symmetric channel state. (d), (e) show the reconstruction results of received signals after applying the proposed DeepPASIC.}
    \vspace{\imagespace}
    \label{fig:image_grid}
\end{figure*}

\vspace{\sectionspace}
\section{Simulation Results and Discussions}

% \vspace{\sectionspace}
\subsection{Dataset and Simulation Settings}
%The ImageNet dataset is a widely used benchmark for image classification tasks. 
The ImageNet2012 dataset\cite{deng2009imagenet} is employed that comprises a training set of 1.2 million images across 1,000 classes and a test set of 50,000 images. We randomly selected 50,000 images from the training set as our training data, and 5,000 images from the test set as our validation set. Each image was first center-cropped to a size of 3$\times$256$\times$256, then converted to a floating-point number with a value range of 0 to 1 to be used as a training sample. Since the encoder conducts two downsampling operations, its output shape is $E\times$64$\times$64, and the semantic encoding length can be controlled by changing the number of semantic feature layers $E$. 

\begin{figure}[t]
\centering
\includegraphics[width=3.3in]{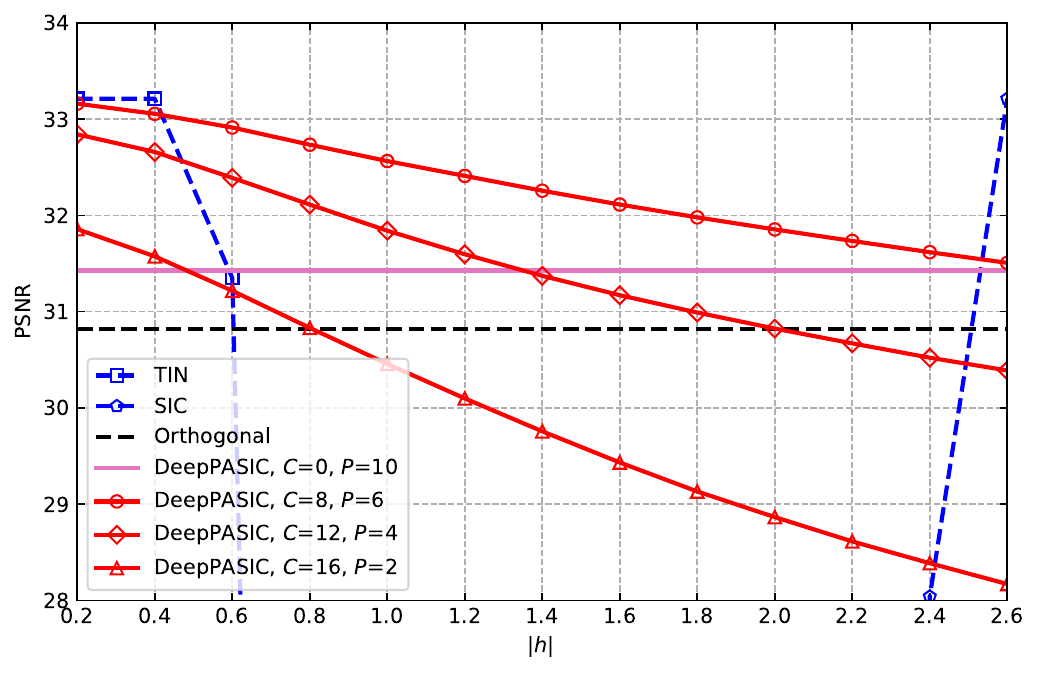}
\caption{PSNR comparisons between different interference management methods with a similar symbol rate.}
\vspace{\imagespace}
\label{fig:psnrvsh}
\end{figure}

\begin{figure}[t]
\centering
\includegraphics[width=3.3in]{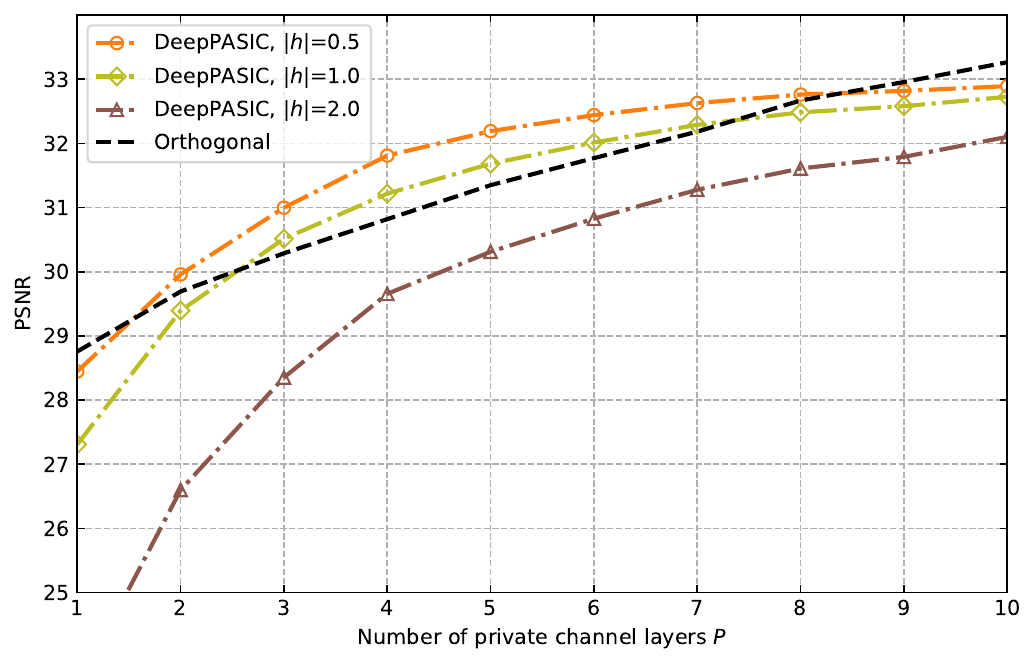}
\caption{PSNR comparisons between DeepPASIC ($E=16$) and the orthogonal bit transmission under varying $|h|$.}
\vspace{\imagespace}
\label{fig:psnrvspr}
\end{figure}

The DeepPASIC model was trained and validated on a two-user AWGN IC. Since the transceiver structure and parameters are shared among different users,  the performance is evaluated for User 1 without loss of generality. The transmit power for each user was assumed to be unity, i.e., $P_1=P_2=1$. The direct channel coefficient for user 1 was normalized to $|h_{11}|=1$, while the interference channel coefficient from user 2 to user 1 was denoted as $h:=h_{21}$ for simplicity. Then the transmit signal-to-noise ratio (SNR) can be calculated as $1/\sigma^2$. The training and testing processes are conducted under interference-constrained scenarios, with the transmit SNR fixed at 15 dB. %For training the autoencoder and the GAN separator in steps 1 and 2, the Adam optimizer is employed with a learning rate of $1 \times 10^{-4}$ and a batch size of 4. The hyperparameters for the training objective of the GAN in (\ref{eq:trainG}) are set to $\alpha=1$ and $\beta=0.01$. Step 3 reduces the learning rate to $5 \times 10^{-5}$. Each step is trained for 12 epochs separately.

To evaluate the performance of DeepPASIC under varying levels of interference, three conventional interference management schemes are employed as baselines: Orthogonal with time division, TIN, and SIC. These methods employ joint photographic experts group (JPEG) source coding, turbo channel coding with a 1/3 code rate, and a 16 quadrature amplitude modulation (QAM) scheme, ensuring identical symbol rates with DeepPASIC for a fair comparison.

\vspace{\sectionspace}
\subsection{Numerical results}
Some visualized results are shown in Fig. \ref{fig:image_grid}. As illustrated in Fig. \ref{fig:image_grid} (d) and (e), although the signal power is equivalent to the interference power, our proposed method can effectively eliminate interference from the superimposed signal. The reconstructed image after decoding exhibits no observable semantic distortion compared to the original image. In contrast, as depicted in Fig. \ref{fig:image_grid} (c), the direct reconstruction of the interference signal without the interference separator and two-stage transmission would result in a severe overlap.

The PSNR performance of DeepPASIC and three baseline approaches for User 1 under varying interference coefficients $|h|$ is illustrated in Fig. \ref{fig:psnrvsh}. We evaluate the following four semantic encoding lengths: $E \in \{10, 14, 16, 18\}$, under the constraint of 20 layers of total orthogonal resources. For a fair comparison, the JPEG source compression rate in the conventional bit communication baseline method is adjusted to ensure that the total number of transmitted symbols is consistent with DeepPASIC.

In Fig \ref{fig:psnrvsh}, the orthogonal scheme is unaffected by interference but requires higher source compression, leading to higher distortion. The TIN scheme performs best in low interference but degrades rapidly as interference increases, as errors exceed the error correction capability. Similarly, the SIC scheme is optimal only for sufficiently high interference. In contrast, the performance of DeepPASIC decreases gradually as $|h|$ increases and outperforms the baselines around the medium interference level.
Besides, it can be observed that the overall PSNR performance of DeepPASIC initially improves and then declines as the encoding length $E$ decreases.
This observation suggests a trade-off between the amount of public part and the semantic encoding length. Increasing the number of public parts introduces more interference. On the other hand, subject to a fixed total resource, more resources can be allocated to improve the encoding length, potentially enhancing the representation capability of the semantic encoder.

Fig. \ref{fig:psnrvspr} illustrates that with a constant semantic encoding length $E=16$, as the number of private channel layers $P$ increases, the PSNR performances of DeepPASIC improve, which can be attributed to the availability of more prompt information to mitigate interference. On the other hand, increasing $P$ necessitates the occupation of more orthogonal resources throughout the transmission process. Therefore, for the orthogonal scheme with the same symbol rate, more bits can be utilized to encode a single image, allowing a lower compression ratio to obtain better image quality.
In scenarios of medium interference ($|h|=1$) and weak interference ($|h|=0.5$), the DeepPASIC with certain $P$ values outperforms the orthogonal bit transmission scheme. For low values of $P$, the performance is inferior to the orthogonal scheme due to severe interference and insufficient capacity for private information. Conversely, for higher values of $P$, the performance also lags behind the orthogonal scheme as the marginal effects in semantic communication are more pronounced than in traditional communication methods.

Although the proposed DeepPASIC exhibits promising results in the moderate interference regime, its performance degrades as the interference intensifies, as shown in Fig. \ref{fig:psnrvsh} and Fig. \ref{fig:psnrvspr}. This limitation could be attributed to the inherent constraints of deep learning models, which may struggle to accurately capture and mitigate strong interference patterns in the input data.

\vspace{\sectionspace}
\section{Conclusions}
In this letter, a multi-user semantic communication system named DeepPASIC has been established to address the challenge of moderate ICs, which is intractable at the traditional syntactic level. By partitioning the semantic features of each user into private and common parts, the private part can serve as a prompt to assist in canceling the interference suffered by the common part at the semantic level. Simulation results demonstrate that the proposed DeepPASIC outperforms conventional interference management strategies under moderate interference conditions. Further research is warranted to address the limitation of semantic interference cancellation under strong interference conditions.

\vspace{\sectionspace}
\bibliographystyle{IEEEtran}
\bibliography{references.bib}

\vspace{\sectionspace}

\end{document}